# Meso-scale phenomena from compromise

*------ a common challenge, not only for chemical engineering*

*Jinghai LI，Wei GE and Mooson KWAUK*

*State Key Laboratory of Multi-phase Complex Systems*

*Institute of Process Engineering, Chinese Academy of Sciences*

*Beijing 100190, People's Republic of China, Email: jhli@home.ipe.ac.cn*

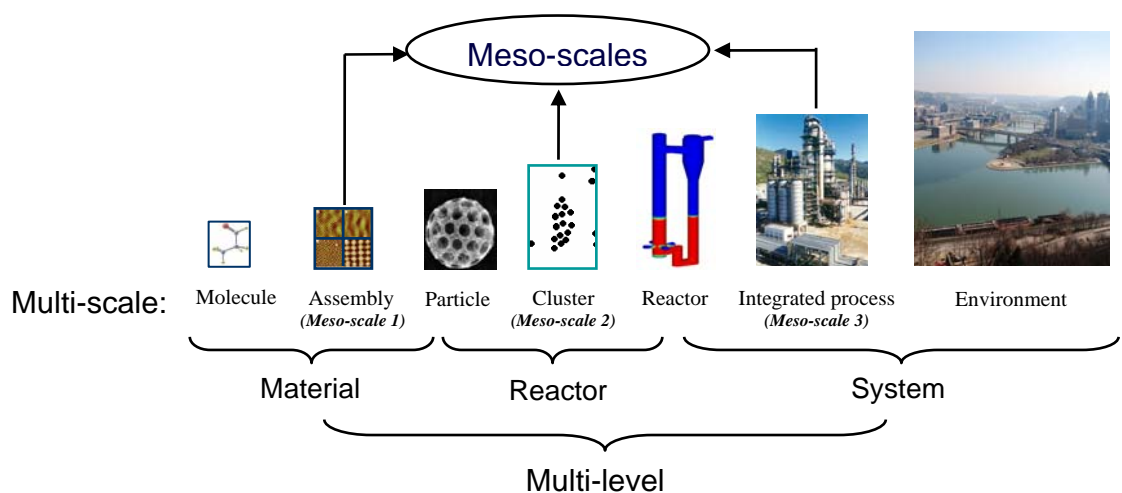

Fig. 1 Challenges at three meso-scales of the multi-level hierarchy of chemical processes

Chemical processes are multi-levelled, consisting of **material, reactor and system,** as shown in Fig. 1, and each level is further featured with its multi-scale structure. The level of **material** covers the domain of research for chemists, which consists of three scales: molecule, molecular assembly and bulk material (e.g. particle, tube or film); the level of **reactor** is the domain of chemical engineers, concerning the scales of particle, particle cluster and reactor; the level of **system** represents the domain of process system engineers and environmentalists, covering single reactor, integrated process and ecological system.

Although these three levels engage in totally different problems, phenomena, and performances at different scales, a **common feature** is that for the much we know about two neigboring scales belonging to each level, we know far less about the meso-scales in-between. For instance, though we have identified the detailed structures of molecules and the properties of materials, we know far less about how to manipulate the meso-scale structure (**the first meso-scale**) to optimize the properties of materials. We comprehend the flow, transport and reaction around a single particle and have experience on the overall performances of various reactors, but know far less on what happens at meso-scale of particle clusters (**the second meso-scale**) and how they influence transport and reaction. At the system level, we have known a lot on the design of a single reactor and the ecological effect of the output of the reactor, but have limited knowledge on how to integrate different reactors and processes (**the third meso-scale**) to realize the *circular economy* to minimize detrimental ecological effects. In conclusion, most chemical processes and materials feature multi-scale structures while the meso-scale phenomena are recognized as the **bottlenecks** in scaling-up processes and in manipulating material structures. What happen at meso-scales, and in particular, the three meso-scales shown in Fig. 1 and discussed above, are believed to be important focii directing future research in the field of chemical engineering. A breakthrough in this respect is likely to lead to significant progress in the field.

**Meso-scale phenomena** represent a challenge not only to chemical engineering but also to the whole spectrum of science and technology. In fact, the term, meso-scale, was used in atmospheric science, referring to the scale of thunderstorm (1) much earlier than in chemical engineering (2) for describing particle clusters. As science continues to expand, downward to the scale of elemental particles and upward to the mega-scale of the universe, people have come to recognize the challenging problems between two neighboring scales for different levels. Researchers have turned their attention to the correlation between micro- and macro-scales, thus touching complexity science (3). In fact, when we were once frustrated by being not able to pinpoint the intrinsic mechanism of the particulate system we were studying -- though we already grasped much of the details at micro-scale and the global nature at macro-scale -- we believed that something critical for the system at the meso-scale must have been missing, thus arousing our attention to this meso-scale. Accordingly, we redirected our efforts in fluid dynamics of multi-phase systems to meso-scale modeling (2) and in turbulence research to large-eddy simulation (4). As shown in Fig. 2, chemists have begun to recognize the importance of phenomena larger than the molecule scale, while chemical engineers started to pay more attention to processes occurring at smaller scales. Therefore, meso-structure became the common ground of our study due to its pivotal influence on the global behaviors of chemical systems. For example, Fig. 2 illustrates that the intrinsic reaction of the atoms and molecules, studied by chemists, is coupled with the transport process, investigated by chemical engineers, at nano-scale to determine the morphology of the material, which presents a compromise of the mechanisms at the meso-scale. Besides chemical science, biology and physics, too, started to recognize the importance of meso-scales. For instance, formulating the sequence of amino-acid residues and the 3-D structure of proteins is not sufficient to reveal the folding and unfolding mechanism at meso-scale and knowing the behavior of electrons and the properties of materials is not sufficient for understanding the mechanism of superconductivity. Breakthrough in understanding meso-scale phenomena is expected to contribute to significant progresses in both science and engineering.

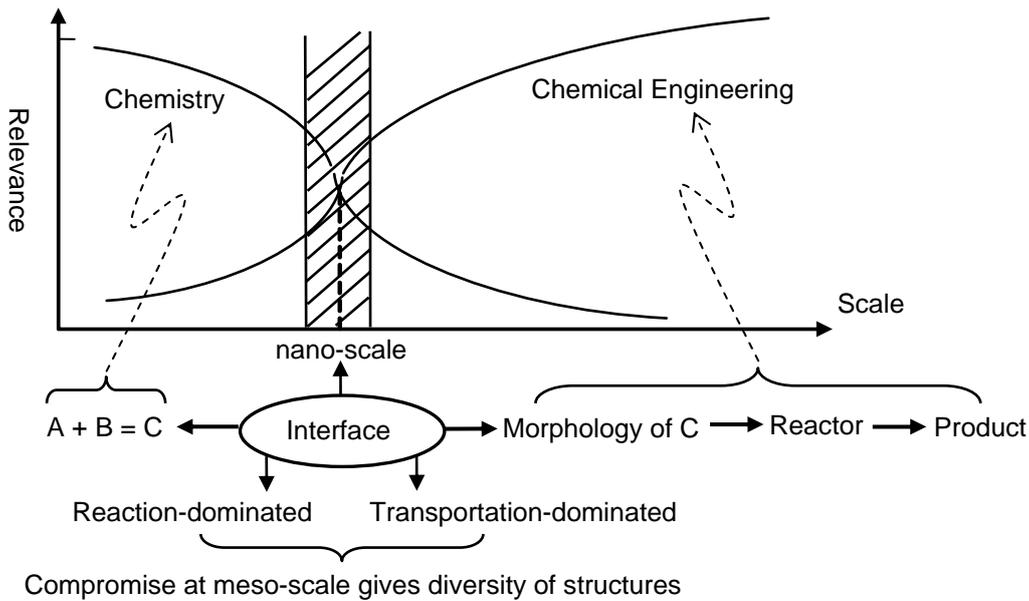

Fig. 2 Interplay of chemistry and chemical engineering

While working on meso-scale phenomena, what first came to our attention was the **dominant mechanism** of the complexity at this scale. Is it independent of adjacent scales? How is it related to micro- and macro- mechanisms? Clarifying these points is crucial. In our study of multi-scale structures, we have come to realize that the complexity of meso-scale phenomena originates from their strong dependence on both micro- and macro-mechanisms, implying that the meso-scale phenomena provide a bridge between micro- and macro-behaviors. For a specified problem, different meso-scale structures define different macro-behaviors, yielding unique flexibility to optimize macro-phenomena by manipulating meso-scale structures. This common nature of complex systems contributes to the diversity of the world, enabling engineers to create new methods, new processes and new materials. Now what is the driving force for the meso-scale phenomena? Micro-mechanisms or macro-dominants? We believe that the compromise between the dominant mechanisms formulatable at the macro-scales is the origin of meso-scale structures (5), though, in addition, subject to micro-mechanisms. Such compromise usually takes place both in space and in time, leading to spatio-temporal dynamic changes. For instance, the compromise between gas behavior and liquid behavior produces bubbles; the compromise between gas behavior and solid particle behavior induces particle clustering; the compromise between viscous behavior and inertial behavior leads to turbulence; two beams of light compromise to display interference patterns. **Compromise**, which exists universally, is crucial to a proper understanding of meso-scale phenomena. Compromise is an intrinsic nature of the world, governed by global mechanisms and subject to micro-mechanisms in every micro-element, thus connecting the micro- and the macro-scales. Mathematically, the dominant mechanisms can, in general, be expressed as extreme tendencies of individual mechanisms, either minimum or maximum, and the compromise can be formulated as mutually constrained conditional extremum between these extreme tendencies (6).

**Holism** purports to comprehend complex systems as a whole, that is, focusing on the global behaviors and corresponding dominant mechanisms (i.e. top-down), while **reductionism** pays more attention to understanding system behaviors at the higher scales by revealing detailed mechanisms at the lower scales (i.e. bottom-up) ----- both of help to better understanding of complex systems. However, considering the joint dominance of both macro-dominants and micro-mechanisms over the meso-scale phenomena, isolated progresses without cross-disciplinary connection, that is, insufficient attention to global behaviors for reductionism or ignorance of micro-mechanisms for holism, might adversely affect the final solution of many complex issues. Integrating the knowledge bases at both ends of the meso-scale helps to establish a pathway to solve engineering problems. Meso-scale is a common ground where both micro-mechanisms and macro-dominants meet, to provide a common focus for theoretical and computational research. We should therefore focus on meso-scale structure by combining our existing knowledge of both micro-scale and macro-scale. This is expected to result in a change in research strategies on complex systems.

Progress in understanding meso-scale phenomena would also lead to advancement in **methodology** and in tools for research, and *vise versa*. Since the computational complexity and cost are likely scale-dependent such that complexity decreases as the cost increases from macro-scale to micro-scale, the computer performance can be optimized by analyzing the distribution of complexity and cost at different scales. For instance, understanding the stability at meso-scale led to the possibility of establishing a new paradigm of multi-scale parallel computation by realizing the structural consistency among the modeling software and hardware, to yield much higher efficiency and lower cost. With such a multi-scale computational system, fully discrete simulation could be accelerated by first performing macro-scale and meso-scale simulation and then followed by discrete simulation at the micro-scale, to significantly shorten the evolution time to the final steady state (7). This means that computer scientists need to work together with engineers, who are familiar with various characteristic structures in engineering, to develop high-efficiency computers. In this regard, meso-scale structures are also a focus due to its common nature and criticality for complex systems. Meso-scale is a joint area for both holism and reductionism. To promote this direction, methodologies in holism and in reductionism need to be integrated to modify the conventional methodologies to formulate new theories and new computation paradigms. That may be the reason why **meso-scale has** become a hot word in recent years Accordingly, validation experiments should be directed also to this scale.

Now the **conclusion** for this commentary. We live in a multi-scale complex world, facing multi-scale problems every day, and we need to follow the multi-scale approach. Among the multiple scales, meso-scale represents the common ground where macro-dominants and micro-mechanisms meet, forming a bridge between the micro and the macro scales. Breakthrough in meso-scale study will lead to significant progresses in engineering science, contributing at the same time to the evolution of research methodology and tools. This is a common challenge not only for chemical engineering.